\documentclass{emulateapj}
\usepackage{apjfonts}
\usepackage{epsfig,color}
\usepackage{mathrsfs}
\usepackage{ulem}

\def\sss{\scriptscriptstyle}
\def\bhm{M_{\bullet}}

\def\rhbeta{R_{\rm H\beta}}

\def\rblrI{R_{_{\rm BLR,I}}}
\def\rblrII{R_{_{\rm BLR,II}}}

\def\CI{C_{_{\rm I}}}
\def\CII{C_{_{\rm II}}}

\def\Mdot{\dot{M}_{\bullet}}

\def\ergs{\rm erg~s^{-1}}

\def\fhbetaI{F_{_{\rm H\beta,I}}}
\def\fhbetaII{F_{_{\rm H\beta,II}}}
\def\fionI{F_{_{\rm ion,I}}}
\def\fionII{F_{_{\rm ion,II}}}  

\def\feii{Fe {\sc ii}}

\def\mathdotM{\dot{\mathscr{M}}}

\def\NH{N_{\rm H}}
\def\oiii{[O\,{\sc iii}]}

\def\rblr{R_{\sss{\rm BLR}}}
\def\rg{r_{\rm g}}

\def\sunm{M_\odot}

\def\sigmaT{\sigma_{\sss{\rm T}}}

\def\taublr{\tau_{_{\rm BLR}}}

\def\tauNIR{\tau_{_{\rm NIR}}}
\def\tauI{\tau_{_{\rm H\beta,I}}}
\def\tauII{\tau_{_{\rm H\beta,II}}}
\def\xin{x_{\rm in}}

\def\xiIA{\xi_{_{\rm IA}}}
\def\xiIB{\xi_{_{\rm IB}}}
\def\xiIIA{\xi_{_{\rm IIA}}}
\def\xiIIB{\xi_{_{\rm IIB}}}

\def\pthetaI{\langle\Theta_{\rm I}\rangle}
\def\pthetaII{\langle\Theta_{\rm II}\rangle}

\def\vecn{\vec{n}}
\def\vecnn{\vec{n}_0}
\def\vecnnn{\vec{n}_1}

\journalinfo{The Astrophysics Journal: Vol. 796:?? (15pp), 2014 December 1 (in press)}
\slugcomment{Received 2014 May 11;  accepted 2014 October 18}

\begin{document}

\title{Self-shadowing Effects of Slim Accretion Disks in Active Galactic Nuclei:\\
Diverse Appearance of the Broad-line Region}

\author
{Jian-Min Wang\altaffilmark{1,2},
Jie Qiu\altaffilmark{1}, 
Pu Du\altaffilmark{1},
and Luis C. Ho\altaffilmark{3,4}
}

\altaffiltext{1}
{Key Laboratory for Particle Astrophysics, Institute of High Energy Physics,
Chinese Academy of Sciences, 19B Yuquan Road, Beijing 100049, China.}

\altaffiltext{2}
{National Astronomical Observatories of China, Chinese Academy of Sciences,
 20A Datun Road, Beijing 100020, China}

\altaffiltext{3}
{Kavli Institute for Astronomy and Astrophysics, Peking University, Beijing 100875, China} 

\altaffiltext{4}
{Department of Astronomy, School of Physics, Peking University, Beijing 100875, China} 

\begin{abstract}
Supermassive black holes in active galactic nuclei (AGNs) 
undergo a wide range of accretion rates, which lead to diversity of appearance.
We consider the effects of anisotropic radiation from 
accretion disks on the broad-line region (BLR), from the 
Shakura-Sunyaev regime to slim disks with super-Eddington accretion rates. 
The geometrically thick funnel of the inner region of slim disks produces
strong self-shadowing effects that lead to very strong anisotropy of the 
radiation field. We demonstrate that the degree of anisotropy of the radiation 
fields grows with increasing accretion rate.  As a result of this anisotropy, 
BLR clouds receive different spectral energy distributions depending on their 
location relative to the disk, resulting in diverse observational appearance of 
the BLR. We show that the self-shadowing of the inner parts of the disk 
naturally produces two dynamically distinct regions of the BLR, depending on 
accretion rate. These two regions manifest themselves as kinematically 
distinct components of the broad H$\beta$ line profile with different line 
widths and fluxes, which jointly account for the Lorentzian profile generally 
observed in narrow-line Seyfert 1 galaxies.  In the time domain, these
two components are expected reverberate with different time lags with respect 
to the varying ionizing continuum, depending on the accretion rate and the 
viewing angle of the observer.  The diverse appearance of the BLR due to 
the anisotropic ionizing energy source can be tested by reverberation mapping 
of H$\beta$ and other broad emission lines (e.g., \feii), providing a new 
tool to diagnose the structure and dynamics of the BLR.  Other observational 
consequences of our model are also explored.
\end{abstract}


\section{Introduction}
Broad Balmer emission lines in active galactic nuclei (AGNs) are main 
features of their optical spectra. As originally suggested by Woltjer (1959), 
the large widths of the emission lines are generally thought to arise from 
virial motions of clouds orbiting around a central supermassive black hole
(Rees 1984; Osterbrock \& Mathews 1986). It has been established that the 
broad emission lines arise from clouds in the broad-line region (BLR) 
photoionized by a central continuum source produced by an accretion disk around
the central black hole (Rees 1984; Ho 2008; Netzer 2013). The smooth profiles 
of AGN spectra suggest that the number of clouds is large (Arav et al. 1997, 
1998).  Reverberation mapping of $\sim$ 50 AGNs has revealed that the size of 
the BLR, as measured for H$\beta$, correlates well with the continuum 
luminosity (Kaspi et al. 2000); the latest study of Bentz et al. (2013) reports
$\rhbeta\approx 33~ L_{44}^{0.53}$ltd, where $\rhbeta$ is the distance of the 
H$\beta$-emitting regions from the center and $L_{44}$ is the continuum 
luminosity at 5100 \AA\, in unit of $10^{44}\ergs$.
Defining the ionisation parameter as
$U=Q_{\rm ion}/4\pi R_{\rm H\beta}^2c n_e$, where $n_e$ is the electron 
density of the clouds, $Q_{\rm ion}$ is the rate of ionizing photons, and $c$ 
is the speed of light, this empirical $\rblr-L$ relation 
can be explained by photoionized BLR clouds that roughly have constant 
$n_eU$ (e.g., Wandel et al. 1999; Bentz et al. 2013). Here $Q_{\rm ion}$ 
($\propto L_{\rm ion}$, the ionising luminosity) is 
proportional to $L_{5100}$ for a canonical spectral 
energy distribution (SED). This relation assumes that all clouds in the 
BLR and the observer at infinity see the same SED of the ionizing source. 
However, we still poorly understand the geometry and dynamics of the BLR.
Only a limited sample of AGNs has been studied with reverberation mapping, 
and the intrinsic scatter of the $\rhbeta-L_{5100}$ relation is poorly 
unknown. The BLR of AGNs shows very diverse spectral properties (e.g., 
Boroson \& Green 1992; Shen \& Ho 2014).  This leads Peterson (2006) to ask: 
``Where does the energy to power the emission lines come from? The BLR energy 
budget problem is still unsolved. The observed AGN continuum is neither 
luminous enough nor hard enough to account for the broad lines. Does the BLR 
see a different continuum than we do, or is there an energy source we have not 
yet recongnized?"

From a theoretical perspective, AGNs with very high accretion rates may offer 
a promising avenue to addressing some of these longstanding problems.  
Accretion disks become geometrically thick with increasing accretion rates so 
that self-shadowing effects are important, leading to strongly anisotropic 
illumination of the BLR.  AGNs with accretion rates in the Shakura \& Sunyaev 
(1973, hereafter SS) regime have been extensively studied.  Netzer (1987) 
discusses the effects of anisotropic emissions from geometrically thin disks 
on emission lines; he treats limb darkening and the 
$\cos\theta(1+2\cos\theta)$ angular dependence of the radiation intensity, 
where $\theta$ is the angle of line of sight to the disk axis. 
Although the radiation from geometrically thin disks shows anisotropy through 
the $\cos\theta$ term, the observational consequences on the BLR are 
minimal.  By contrast, accretion flows with high accretion rates, known as 
slim disks (Abramowicz et al. 1988), develop sharp funnels as a consequence 
of their geometrically thick structure when they have accretion rates higher 
than a few $L_{\rm Edd}/c^2$, where $L_{\rm Edd}$ is the Eddington luminosity. 
This is expected to produce much more dramatic anisotropic illumination of the 
BLR.  In the nearby Universe, slim disks are believed to power narrow-line 
Seyfert 1 galaxies (NLS1; Osterbrock \& Pogge 1985), which contain relatively 
low-mass black holes with higher accretion rates (Boller et al. 1996; Wang \& 
Netzer 2003). Direct observational evidence for slim disks is still relatively 
scarce (but see Kawaguchi et al. 2004; Desroches et al. 2009; Jin et al. 2009), 
limited by the availability of reliable SED predictions for this regime of 
accretion rates. Some effects of the anisotropic radiation on SEDs have been 
discussed by Czerny \& Elvis (1987) for standard disks with higher 
accretion rates and Madau (1988) for simplified tori. Watarai et al. (2005) 
and Li et al. (2010) studied the geometric effects of slim disks on SEDs of 
ultraluminous X-ray sources.  However, the effects of anisotropic illumination 
of a slim disk on the BLR of AGNs have not be explored in the
published literature.  This is the subject of our current work.

Beginning with a general treatment of the structure of accretion disks from 
low to high accretion rates, we show that the self-shadowing effects of the
funnel in a slim disk produce a strong angular dependence of the radiation 
field. The high degree of anisotropy of the radiation field naturally 
defines two spatially and kinematically distinct regions in the BLR.  \S 2 is 
devoted to the details of the anisotropic radiation field.  \S 3 elucidates the 
two-component nature of the H$\beta$-emitting region and relevant consequences 
of the BLR in the time domain.  Observational implications beyond the BLR are 
discussed in \S4.  The last section presents a brief summary. 

\section{Anisotropic Radiation of Slim Accretion Disks}
\subsection{Slim Disks} 
The geometry of accretion disks, which is controlled by the accretion mode,  
determines the anisotropy of the radiation field.  The accretion mode is 
governed by the dimensionless accretion rate, defined by
\begin{equation}
\mathdotM=\frac{\Mdot}{L_{\rm Edd}c^{-2}},
\end{equation}
where $\Mdot$ is the accretion rate, the Eddington luminosity 
$L_{\rm Edd}=4\pi G\bhm cm_p/\sigmaT$, 
$\bhm$ is the black hole mass, $G$ is the gravitational constant, $m_p$ 
is the proton mass, and $\sigmaT$ is the Thomson cross section. SS disks
in the sub-Eddington regime ($\mathdotM\lesssim 1$) are characterized by the
following features (see Frank et al. 2002): (1) Keplerian angular momentum 
distribution; (2) energy balance between dissipation and radiative cooling 
is localized so that radial advection of energy can be neglected; and (3) 
the disk maintains a geometrically thin configuration. The anisotropy of an 
SS disk simply depends on the factor $\cos \theta(1+2\cos\theta)$ because 
the disk is geometrically thin. It should be noted that most AGNs studied 
with reverberation mapping (Bentz et al. 2013) are in the SS regime. The 
anisotropy of SS disks is much weaker than that of slim disks.

When the accretion rate $\mathdotM$ increases, the above-mentioned features 
dramatically change. In such a situation, the angular momentum distribution 
is sub-Keplerian, the radial velocity of the accretion flow becomes comparable 
with the Keplerian rotation, which leads to strong photon trapping and a 
saturated luminosity, and, in particular, the inner parts of the disk develop 
a sharp funnel.  The funnel produces strong self-shadowing effects depending 
on the actual accretion rates. There has been increasing attention on 
numerical simulations of slim disks (Ohsuga et al. 2005; McKinney et al. 2014; 
Sadowski et al. 2014; Yang et al. 2014; Jiang et al. 2014). 
The general properties of slim disks from these simulations agree with 
the solutions derived from the vertically averaged equations of the disk. 

For explicit discussion of the effect of slim disks on the BLR, we start from 
the averaged equations for slim disks. Hoshi (1977) introduced the polytropic 
relation $p=k_0\rho^{1+1/N}$, where $N$ is the polytropic index and $k_0$ is
a constant, and vertically integrated the equations for an axisymmetric accretion 
disk. For convenience to the readers, we give the 
vertically averaged equations of slim disks (e.g., Muchotrzeb \& Paczynski 
1982; Matsumoto et al. 1984; Abramowicz et al. 1988; Watarai et al. 2005). 
In this paper, we use the pseudo-Newtonian potential given by 
$\Phi=-G\bhm/(R-2\rg)$, where $R=(r^2+z^2)^{1/2}$ and
$\rg=G\bhm/c^2$ is the Schwartzschild radius.

\begin{figure*}[t!]
\begin{center}
\includegraphics[angle=-90,width=0.65\textwidth]{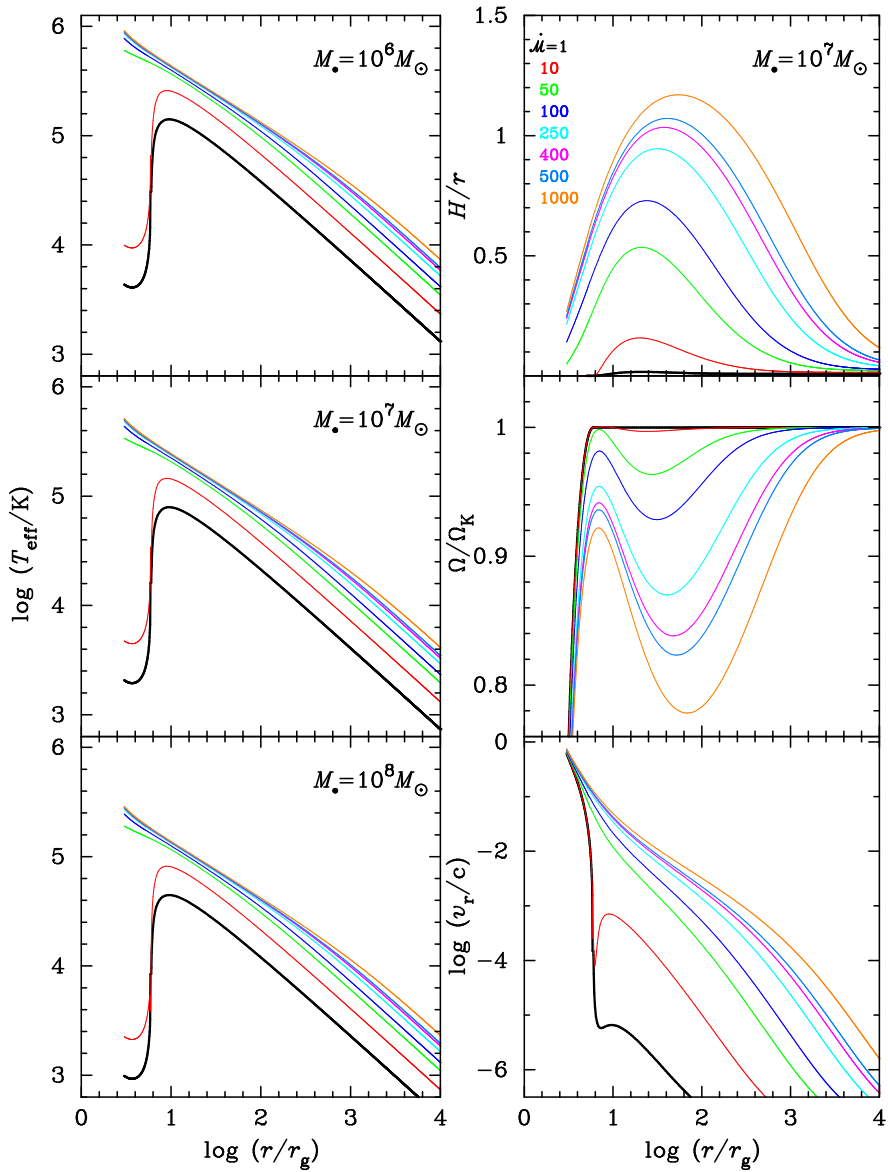}
\end{center}
\caption{\footnotesize Structure of slim disks for given black hole masses 
and accretion rates (labeled by lines of different color). The panels 
on the left column plot the radial variation of effective temperature 
for $\bhm=10^6,10^7,10^8 \,\sunm$.  The panels on the right column plot the 
radial variation of $H/r$, $\Omega/\Omega_{\rm K}$, and $v_{\rm r}/c$ only for 
$\bhm=10^7 \,\sunm$, as these quantities are rather insensitive to $\bhm$. 
Slim disks are characterized by (1) $H/r\sim 1$, (2) effective
temperature $T_{\rm eff}\propto r^{-1/2}$ and decreases with black hole mass, 
(3) sub-Keplerian rotation, and (4) transonic accretion flows. These properties 
differ from those of Shakura-Sunyaev disks.
}
\label{fig_1}
\end{figure*}
\vglue 0.5cm

\subsubsection{Equations of Slim Disks}
The mass conservation of a stationary slim disk reads 
\begin{equation}
\Mdot=-2\pi r \Sigma v_r, 
\end{equation}
where $v_r$ is the radial velocity of the accreting gas, $\Sigma=2I_{_N}H\rho$ 
is the surface density, $I_{_N}=(2^NN!)^2/(2N+1)!$ is the average coefficient
and $\rho$ is density. Angular momentum conservation is governed by 
\begin{equation}
\Mdot(\ell-\ell_{\rm in})=-2\pi r^2 T_{r\phi}, 
\end{equation}
where $\ell$ and $\ell_{\rm in}$ are the
specific angular momentum at $r$ and at the inner edge $r_{\rm in}$, 
and $T_{r\phi}$ is the turbulent viscosity. The radial motion follows from
\begin{equation}
v_r\frac{dv_r}{dr}+\frac{1}{\Sigma}\frac{d\Pi}{dr}
=\frac{\ell^2-\ell_{\rm K}^2}{r^3}-\frac{\Pi}{\Sigma}\frac{d\ln \Omega_{\rm K}}{dr},
\end{equation}
where $\Pi=\int pdz=2I_{_{N+1}}pH$ is the integrated pressure, $p$ is the total pressure,
$\Omega_{\rm K}=(G\bhm/r)^{1/2}/(r-2\rg)$ is the Keplerian rotation velocity and 
$\ell_{\rm K}=r^2\Omega_{\rm K}$ is the Keplerian angular 
momentum. We adopt the $\alpha$ prescription for the viscosity (SS), such that
$T_{r\phi}=-\alpha \Pi$. The vertical structure maintains a statistic equilibrium 
\begin{equation}
H^2=\frac{2(N+1)I_{_N}}{I_{_{N+1}}}\frac{\Pi}{\Omega_{\rm K}^2\Sigma}, 
\end{equation}
where $H$ is the height of the disk. We use $N=3$ in our calculations.
The non-local energy budget is given by
\begin{equation}
Q_{\rm dis}=Q_{\rm adv}+Q_{\rm rad},
\end{equation}
where $Q_{\rm dis}$ is the rate through which viscosity dissipates the 
gravitational energy of the accretion flow, $Q_{\rm adv}$ is the advection 
rate, and $Q_{\rm rad}$ is the radiative flux from the surface of the disk. 
The dissipation rate is given by
\begin{equation}
Q_{\rm dis}=rT_{r\phi}\frac{d\Omega}{dr}.
\end{equation}
According to the thermal dynamics of the gas, the advection rate is
\begin{equation}
Q_{\rm adv}=\frac{9}{16\pi r}\Mdot T\left(\frac{dS}{dr}\right), 
\end{equation}
where the entropy is given by 
$dS=\left[(12-10.5\beta)d\ln T-(4-3\beta)d\ln \rho\right]$,
$\beta=p_{\rm gas}/(p_{\rm gas}+p_{\rm rad})$ is the fraction of the gas 
pressure relative to the total pressure, $p_{\rm rad}=aT^4/3$ is the 
radiation pressure, and $a$ is the blackbody radiation constant. Using the 
diffusion approximation, the cooling rate can be written as
\begin{equation}
Q_{\rm rad}=\frac{8}{3}\frac{acT^4}{\tau}, 
\end{equation}
where $\tau=(\kappa_{\rm es}+\bar{\kappa}_{_{\rm ff}})\Sigma$ is the optical 
depth. Here $\kappa_{\rm es}=0.34$ is the opacity of electron scattering and the
Rossland mean opacity $\bar{\kappa}_{_{\rm ff}}=6.5\times 10^{22}\rho T^{-3.5}$.
The diffusion approximation holds well for high accretion rates.

\subsubsection{Structure and Radiation}
The detailed structure of slim disks has been presented by Abramowicz et al. 
(1988), Watarai et al. (2005), and Li et al. (2010) for stellar mass black holes 
and by Szuszkiewicz et al. (1996), Mineshige et al. (2000), and Chen \& Wang (2004) 
for supermassive black holes. Here we calculate the structure of slim disks to 
study their self-shadowing effects for supermassive black holes spanning a 
large range of accretion rates.

\begin{figure*}[t!]
\begin{center}
\includegraphics[angle=-90,width=0.9\textwidth]{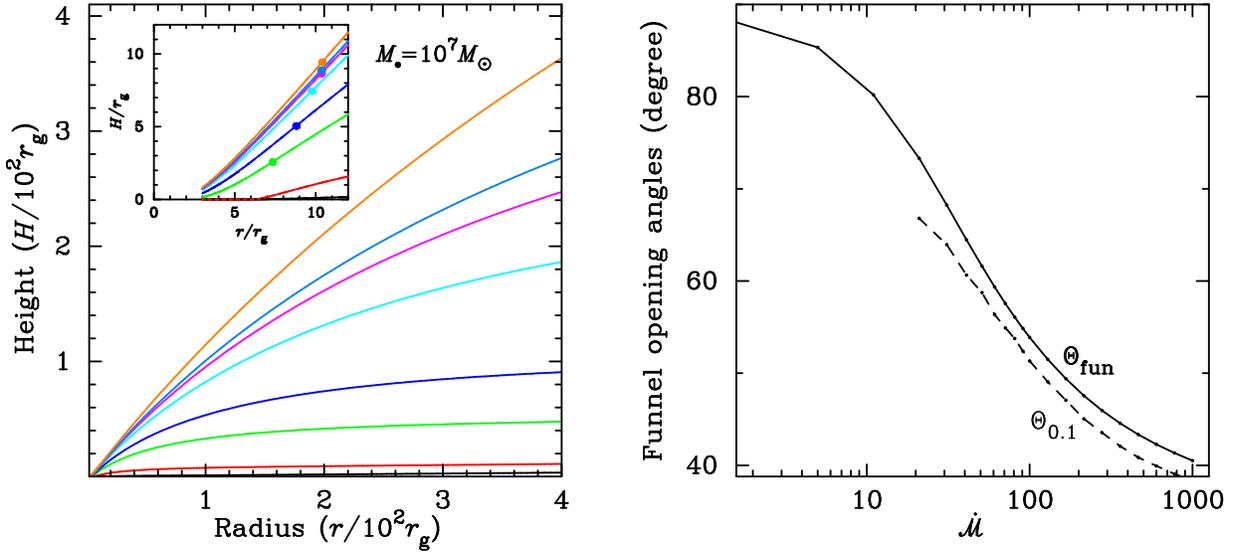}
\end{center}
\caption{\footnotesize ({\it Left}) The variation of the height of slim 
accretion disks with different parameters. We assume a viscosity parameter 
$\alpha=0.1$ and $\bhm=10^7\sunm$.  The values of black hole mass and accretion 
rate are labeled. Slim disks show funnels. In the inserted panel, the points on 
the curves mark the radius ($r_{0.1}$) obtained from Wien's law with
a temperature of 0.1 keV.  For a given accretion rate, the shape of a 
slim disk is very insensitive to black hole mass, but it depends strongly on 
$\mathdotM$ for a given black hole mass. ({\it Right}) Variation of the half 
opening angle of the funnel ($\Theta_{\rm fun}$) as a function of $\mathdotM$; 
here we use $\bhm=10^7\,\sunm$. We find that the shape of the disk at radius $r_{0.1}$ 
$\Theta_{0.1}\lesssim \Theta_{\rm fun}$ with very little dependence on black 
hole mass. }
\label{fig_2}
\end{figure*}
\vglue 0.5cm

Figure 1 shows the disk structure for different black hole masses and accretion 
rates. For SS disks, we still get transonic structures, which 
allow us to avoid the density singularity at the inner edge. The disks are 
generally geometrically thin, namely, $h\lesssim 10^{-2}$, where $h=H/r$,
even in the radiation pressure-dominated regions, have Keplerian angular momentum 
distributions, and effective temperature distribution of 
$T_{\rm eff}\propto r^{-3/4}$. Compared with geometrically thin disks, 
slim disks have the following main dynamical features: (1) $h\sim 1$ for 
$r<10^3\,\rg$, reaching $h \sim 10^{-2}$ in the outer regions; (2) 
the effective temperature $T_{\rm eff}\propto r^{-1/2}$, which is different 
from $T_{\rm eff}\propto r^{-3/4}$ in geometrically thin disks; (3) radial 
velocity of the accretion flow is transonic, $v_r\sim v_{_{\rm K}}$, 
where $v_{_{\rm K}}$ is the Keplerian rotation velocity, leading to fast 
transportation of radiation into the black hole through photon trapping; 
(4) rotation of the accretion flow is sub-Keplerian. These properties 
are generally consistent with Abramowicz et al. (1988), Szuszkiewicz et 
al. (1996), Chen \& Wang (2004), and Watarai et al. (2005). These dynamical 
behaviors lead to radiation properties that differ from those in a SS disk.

The fast radial transportation of accretion flows leads to inefficient 
radiation from the disk surface caused by very large optical depth to Thomson 
scattering. This crucial property has two consequences.  First, as a result of 
the effective temperature dependence of $T_{\rm eff}\propto r^{-1/2}$, the 
emergent spectrum is characterized by $F_{\nu}\propto \nu^{-1}$.  And second, 
the radiative efficiency $\eta\propto \mathdotM^{-1}$, such that the 
luminosity saturates as
\begin{equation}
L_{\bullet}\approx 2L_{\rm Edd}\left[1+\ln \left(\mathdotM/50\right)\right].
\end{equation}
This approximation derives from a fit to numerical results (Mineshige et al. 
2000) and is consistent 
with the self-similar solution for slim disks (Wang \& Zhou 1999). 
The saturated luminosity is only logarithmically dependent on accretion 
rate and is linearly proportional to black hole mass. This unique property 
of slim disks suggests that super-Eddington accreting massive black holes 
(SEAMBHs) potentially may serve as standard candles for cosmology, provided 
that black hole masses can be measured accurately by reverberation mapping 
(Wang et al. 2013).  A recent AGN reverberation mapping campaign shows that 
SEAMBHs are a promising new tool for cosmology of the high-$z$ Universe 
(Du et al. 2014; Wang et al. 2014).

In our simplified model of slim disks, the saturated luminosity still depends 
on $\ln \mathdotM$.  For more accurate application of SEAMBHs as standard 
candles for precision cosmology, it is still important for us to try to 
constrain $\mathdotM$.  Although the emergent spectra is highly insensitive to 
the accretion rate, fortunately, the {\it shape}\ of the slim disk itself 
{\it is}\ sensitive to $\mathdotM$ once accretion rates are sufficiently high. 
As we discuss below, self-shadowing effects arising from the geometric shape of 
the slim disk strongly influence the illumination of the BLR. This gives an 
opportunity to constrain the accretion rate from observations of the BLR 
emission.

\begin{figure*}[t!]
\begin{center}
\includegraphics[angle=-90,width=0.75\textwidth]{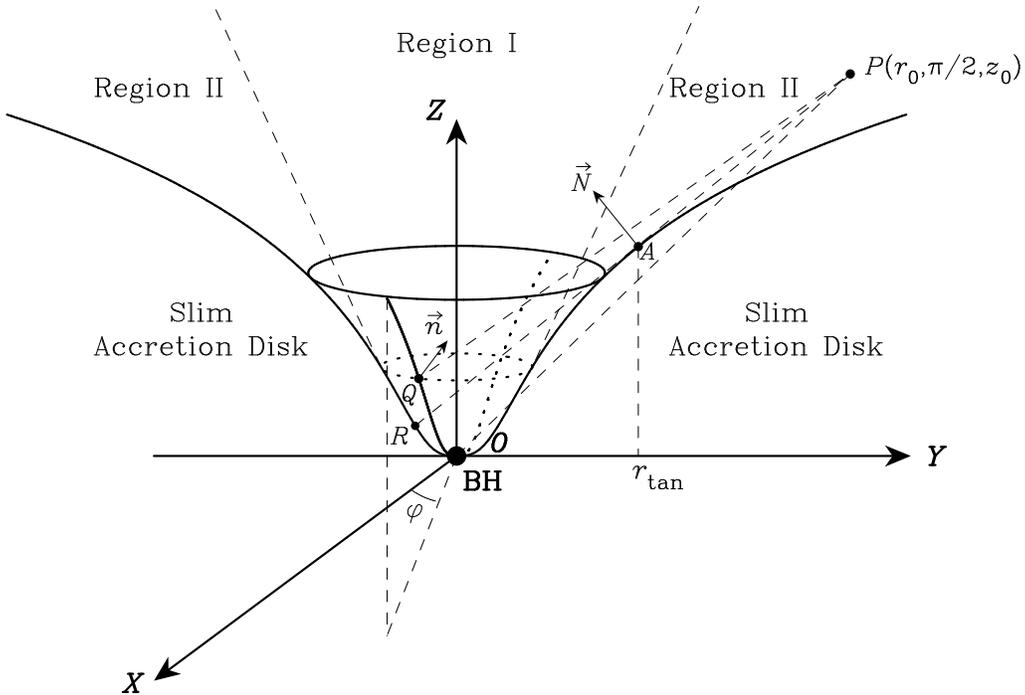}
\end{center}
\caption{\footnotesize Cross section of the system along the $z-$axis of 
symmetry, including the accretion disk and the BLR. The mid plane of the disk 
is in the $XOY$ plane with the origin centered on the black hole. The funnel
of the slim 
disk divides the BLR into two regions: Region I is illuminated by the entire
disk and Region II is illuminated by part of the disk. The point 
$P(r_0,\pi/2,z_0)$ in the BLR has the tangent line $AP$ at point $A$ on the 
surface of the disk and cross point $R$ at the surface of the disk. The direction 
vector of the cloud is $\vecnn$, and the vector normal to line $QP$ is 
$\vecnnn$. }
\label{fig_3}
\end{figure*}
\vglue 0.5cm

\subsection{Geometry and Anisotropic Radiation}
\subsubsection{Funnels in Slim Disks}
We obtain solutions for the structure of accretion disks from sub-Eddington 
accretion rates to extremely high, super-Eddington rates ($\mathdotM\gg 1$). 
This allows us to build up the smooth 
transition from SS to slim disks. The general features of the disk geometry 
are characterized by the appearance of sharper funnels with increasing 
$\mathdotM$. The first row of panels in Figure 1 shows the relative height of 
the disk for different values of $\bhm$ and $\mathdotM$. We find that $h$ is 
very insensitive to the black hole mass. The shape of a slim disk has three
notable features: (1) a funnel develops in the innermost region
[$dh/dr>0$]; (2) a flattened part [$dh/dr<0$ and $h\sim 1$]; 
(3) and a geometrically thin part ($h\sim 10^{-2}$), approaching the SS 
regime, in which the funnel disappears. 

The left panel of Figure 2 shows the height of the disk as a function of 
radius; the funnel structure is clearly evident. The height of the funnel 
increases with accretion rates. There is a critical point, which can be used 
to define the funnel radius ($r_{\rm fun}$), wherein $h$ reachs its maximum 
value [$dh/dr=0$]; we denote the disk height $H_{\rm fun}$ at 
$r_{\rm fun}$.  We define 
$\Theta_{\rm fun}=\arctan\left(r_{\rm fun}/H_{\rm fun}\right)$ as the half 
angle of the funnel. From the numerical results in Figure 1, we approximate 
$r_{\rm fun}/\rg\approx 24\mathdotM_{100}^{0.38}$, where 
$\mathdotM_{100}=\mathdotM/100$. We note that the photon-trapping radius can be
approximated by $r_{\rm ph}/\rg\approx 50(\mathdotM/50)$ (Equation 22 in 
Wang \& Zhou 1999), which is larger than $r_{\rm fun}$ and much more sensitive 
to accretion rate. This guarantees that photon-trapping processes dominate in 
the funnel region and that the BLR size is mainly determined by the saturated 
luminosity.

We digress for a moment to point out the radiation distribution along
with accretion disks. Since the surface flux 
$F\propto x^{-3}\left[1-\left(\xin/x\right)^{1/2}\right]$, the fraction of the 
dissipated energy within $x$ is given by $f(x)=1-3\xin x^{-1}+2\left(\xin x^{-1}\right)^{3/2}$,
where $x=r/\rg$ and $\xin$ is the inner edge of the disk. For a Schwartzschild
black hole, $\xin=6$, and we have $f(x)=(0.5,0.8)$ for $x\approx (24,73)$, namely that
most of the gravitational energy is released within a few tens of Schwartzschild radii.  
Although its formation mechanism remains uncertain,
the corona is expected to follow the distribution of gravitational energy release (e.g., 
Svensson \& Zdzarski 1994; Merloni \& Fabian 2002; 
Wang et al. 2004; Uzdensky 2013). The corona itself may be less anisotropic than the disk,
but it should still be shadowed to the BLR clouds by the slim disk. And while the corona 
generates hard X-rays, the fraction of gravitational energy dissipated in the hot corona 
decreases with increasing accretion rate, generally limited to less than $\sim 15\%$ in 
NLS1s (Wang et al. 2004; Yang et al. 2007; see Table 6 in Grupe et al. 2010). Therefore, 
hard X-rays play only a minor role in powering H$\beta$ emission, compared to UV and soft 
X-rays, which come from the innermost regions of the slim disk. As we show below, 
strong soft X-ray emission is one of the characteristics of slim disks, and the UV and soft 
X-rays are highly anisotropic with respect to the BLR clouds.

The portion of the ionizing source that reaches the BLR is determined by the 
relative orientation between the clouds and the surface of the disk. For 
H$\beta$ line emission, we define the radius of the disk with a temperature 
of $\sim 0.1$ keV (the main drivers of H$\beta$ emissions) 
as $r_{0.1}$. We find $r_{0.1}/\rg\approx 8.0\mathdotM_{100}^{0.18}$.
If the $r\lesssim r_{0.1}$ regions are partially obscured by 
self-shadowing, the region with ionized hydrogen atoms will shrink. It is 
thus useful to define the opening angle of the funnel for the H$\beta$-emitting
region by the shape of the disk at radius $r_{0.1}$.  The right panel of 
Figure 2 shows that $\Theta_{\rm fun}$ and $\Theta_{0.1}$ 
are approximately equal to each other. Consequently, the opening 
angle of the funnel can be determined by the tangent direction at $r_{0.1}$:
\begin{equation}\begin{array}{lll}
\Theta_{0.1}&=&\arctan\left(H_r^{-1}\right)_{0.1} \\
             & &                                    \\
             &\approx & \left\{\begin{array}{ll}
 60^{\circ}-33^{\circ}\log(\mathdotM/50)&{\rm (for~~10\lesssim \mathdotM\lesssim 100)},\\
                 & \\
 52^{\circ}-12^{\circ}\log(\mathdotM/100)&{\rm (for~~\mathdotM\gtrsim 100)},
                \end{array}\right.
\end{array}
\end{equation}
where $H_r=dH/dr$ and the subscript $0.1$ is $H_r$ at $r_{0.1}$. The opening angle 
decreases faster for lower accretion rates ($\mathdotM\lesssim 100$), while it 
changes slower for higher accretion rates (tends to a minimum opening angle for 
cases of very high accretion rates). 

Numerical results show that the funnels are insensitive to black hole mass. In 
this paper, we take $\bhm=10^7\,\sunm$ for most of our calculations.

\begin{figure*}[t!]
\begin{center}
\includegraphics[angle=-90,width=0.5\textwidth]{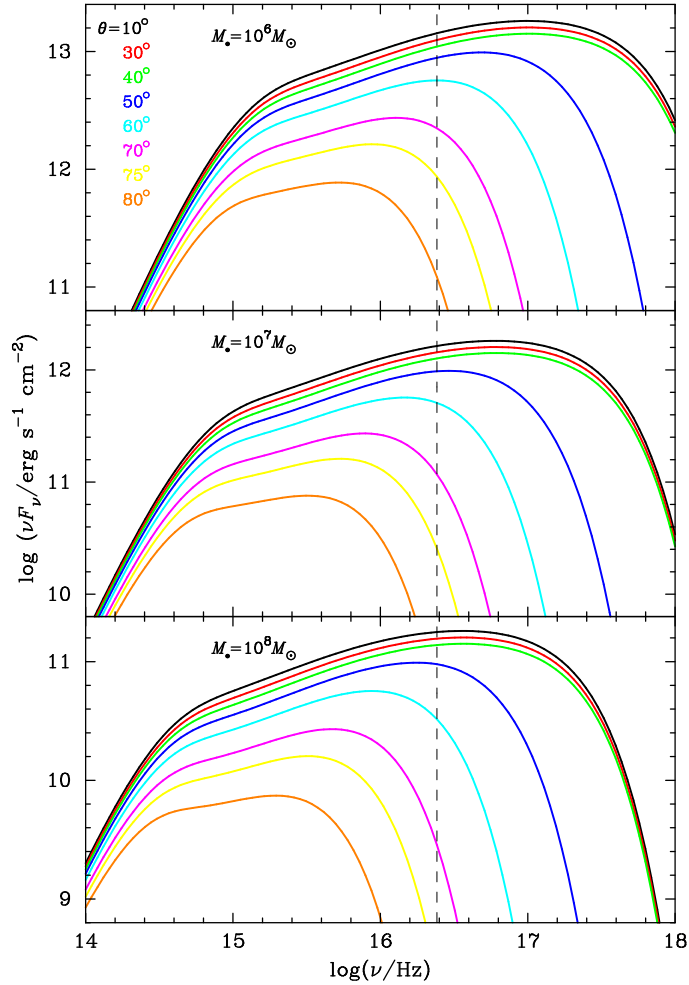}
\end{center}
\caption{\footnotesize Spectral energy distributions of slim disks 
received by BLR clouds at different angles illustrate the anisotropy of 
the radiation field from slim disks. Colored lines correspond to the values of 
parameters labeled with the same color. The orientation angle $\theta$ of one 
cloud in the BLR is defined by $\tan\theta=r_0/z_0$ with respect to the $Z$-axis. 
The received SED strongly depends on $\theta$.  We assume $\mathdotM=500$, and 
the distances of the clouds are assumed to be $r_0=10^4\rg$. 
There is a clear break frequency in the SED due to photon 
trapping; it depends on the black hole mass and viewing angle. An observer at 
high inclinations only receives emission from the parts of the disk without 
photon trapping, so that the break disappears gradually. The dashed line
is 0.1 keV, above which the SED mainly drives H$\beta$ emission.}
\label{fig_4}
\end{figure*}
\vglue 0.5cm

\begin{figure*}[t!]
\begin{center}
\includegraphics[angle=-90,width=0.7\textwidth]{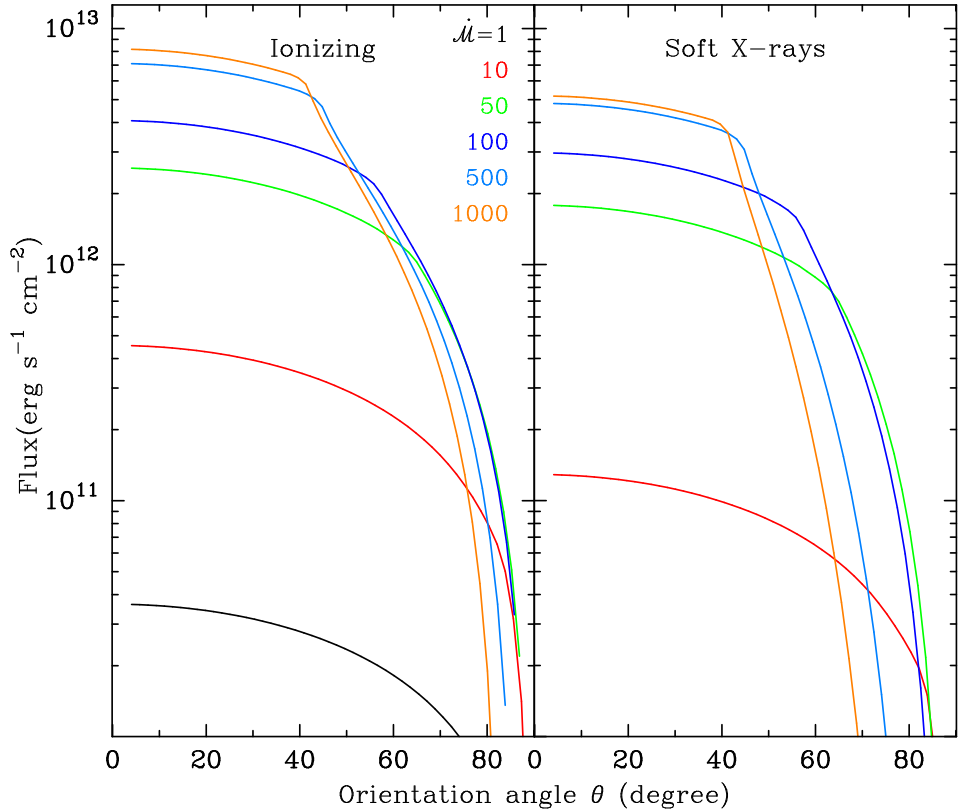}
\end{center}
\caption{\footnotesize  The ({\it left}) ionizing (photons with energy
$\ge 13.6$ eV) flux and ({\it right}) soft X-ray flux of a slim disk as viewed 
by BLR clouds at different angles. We find that the flux decreases sharply beyond 
the half opening angle of the disk funnel.  The anisotropy of emissions received 
by the clouds depends on the orientation angle and accretion rate of the slim disk.  
Comparing the two panels, we find that soft X-rays are more anisotropic
than ionizing (UV + soft X-ray) photons.}
\label{hubble_diagram}
\end{figure*}
\vglue 0.5cm

\subsubsection{Anisotropic Radiation Field}
Canonical spectra of slim accretion disks have been calculated by Szuszkiewicz 
et al. (1996), Wang et al. (1999), Mineshige et al. (2000), and Watarai et al. 
(2005), among others. Previous models approximate the slim disk 
as parallel layers, and  neither the influence of the disk geometry on 
the emergent spectra nor the ionizing spectra received by the BLR clouds has 
been investigated.  The BLR clouds cannot be regarded as a receiver at 
infinity.  Figure 3 illustrates the present scheme for calculating the 
self-shadowing effects on the clouds. We follow the approach used by 
Pacharintanakul \& Katz (1980).  Electron scattering has significant effects 
on the spectra because the Thomson scattering opacity dominates the free-free 
absorption so that the radiation can be modified as grey blackbody radiation 
(e.g., Czerny \& Elvis 1987). In this paper, we include the effects of 
electron scattering on the spectra by simplifying it as modified blackbody 
radiation from the disk surface (e.g., Szuszkiewicz et al.  1996). The 
intensity of radiation from the disk surface is given by
\begin{equation}
I_{\nu}=\frac{2B_{\nu}(T)}{1+\sqrt{1+\kappa_{\rm es}/\kappa_{_{\rm ff}}(\nu)}},
\end{equation}
where $B_{\nu}(T)$ is the Planck function and the free-free opacity is
\begin{equation}
\kappa_{_{\rm ff}}(\nu)=1.5\times 10^{25}\rho T^{-3.5}x^{-3}(1-e^{-x}),
\end{equation}
where $x=h\nu/kT$, $h$ is the Planck constant, $k$ is the Boltzman constant,
and the Gaunt factor is taken to be unity. Here, radiation from the surface 
is isotropic, such that $I_{\nu}$ is independent of angle. 
The flux received by a cloud at point $P(r_0,\pi/2,z_0)$ in the $YOZ$ plane
is given by
\begin{equation}
F_{\nu}(r_0,z_0)=\int_{\mathscr{S}_0}I_{\nu}(r,\phi,z)(\vecn\cdot\vecnnn)(\vecnnn\cdot\vecnn)
                 \frac{d\mathscr{S}}{d_0^2},
\end{equation}
where $d_0$ is the distance of the cloud to the center, $\vecn_0$ is the 
direction of the line-of-sight of the cloud in the BLR, $\vecn$ is the outward 
normal to the surface area element $d\mathscr{S}=r(1+H_r^2)^{1/2}drd\phi$,
$\mathscr{S}_0$ is the area unobscured by the disk itself, and $\vecn_1$ is 
the normal vector from any point $P(r,\phi,z)$ of the disk surface. Appendix 
gives details. It should be noted here that $d_0$ is not much larger than the 
disk size, and the shape of the disk is important to calculate the ionizing 
flux of the BLR clouds. 

Figure 4 shows SEDs received by clouds in the BLR with different orientation 
to the disk. Clouds with $R_0=10^4\rg$ for different viewing angles 
$\theta=\tan^{-1}(r_0/z_0)$ are considered for slim disks with $\mathdotM=500$ 
and $\bhm=10^6,10^7$, and $10^8\,\sunm$. It clearly shows the anisotropy of 
the radiation field. Two effects are seen: (1) The flux received by the clouds 
dramatically decreases with orientation angle by a factor of 30 for 
$\theta=10^{\circ}$ to $80^{\circ}$, which is much steeper than $\cos\theta$; 
(2) the SEDs are significantly softened by self-shadowing at lower latitudes, 
resulting in the lack of photoionizing photons for emission lines. As a 
consequence, the clouds are exposed to SEDs and luminosities different from 
the observer. These two effects may introduce scatter to the $\taublr-L$ 
relation and may allow an opportunity to relax the energy budget in the BLR. 
Self-shadowing effects are very crucial for the manner in which BLR clouds 
respond to the varying continuum.

The theoretical SEDs of slim disks in the optical to UV regime show a 
characteristic form $F_{\nu}\propto \nu^{\alpha_{\nu}}$, with $\alpha_{\nu} 
\approx -1$ within the photon trapping radius, spanning a frequency range 
that depends on accretion rate.  Beyond the photon trapping radius, the disk 
emission approximately follows the standard $\alpha_{\nu} = +1/3$ shape of 
an SS disk.  The transition frequency between these two regimes (from 
$\alpha_{\nu} = +1/3$ to $-1$) depends on $\bhm$, $\mathdotM$, and even BH 
spin.  Neither of these naive expectations compares favorably with 
observations, although the mismatch for SS disks is a well-known problem 
(e.g., Koratkar \& Blaes 1999).  Studies of the composite optical-UV SEDs of 
quasars and AGNs typically find $\alpha_{\nu} \approx -0.3$ to $-0.6$ for 
$\lambda > 1200$ \AA, steepening to $\alpha_{\nu} = -1.4 $ to $-2$ for 
$\lambda < 1000$ \AA\ (e.g., Zheng et al. 1997; Telfer et al. 2002).  The 
average FUV-EUV slope of the 22 sources recently analyzed by Shull et al. 
(2012) is $\langle\alpha_{\nu}\rangle=-0.96\pm 0.41$, with considerable 
variation from object to object.  This observed slope superficially resembles 
the characteristically flat spectrum of a slim disk, but we caution 
against overinterpretation.  More sophisticated calculations are needed to 
produce robust SED predictions for slim disks.  Moreover, we do not have 
complete information on the Eddington ratios of the objects in Shull et al.'s 
sample.  We emphasize, however, that none of the conclusion of this study 
depends on knowledge of the detailed SED of a slim disk.  Instead, we 
simply utilize the fact that, due to self-shadowing and the resulting 
anisotropy, the overall SED exhibits a strong angular dependence, as 
illustrated in Fig 4.

\begin{figure*}[t!]
\begin{center}
\includegraphics[angle=-90,width=0.7\textwidth]{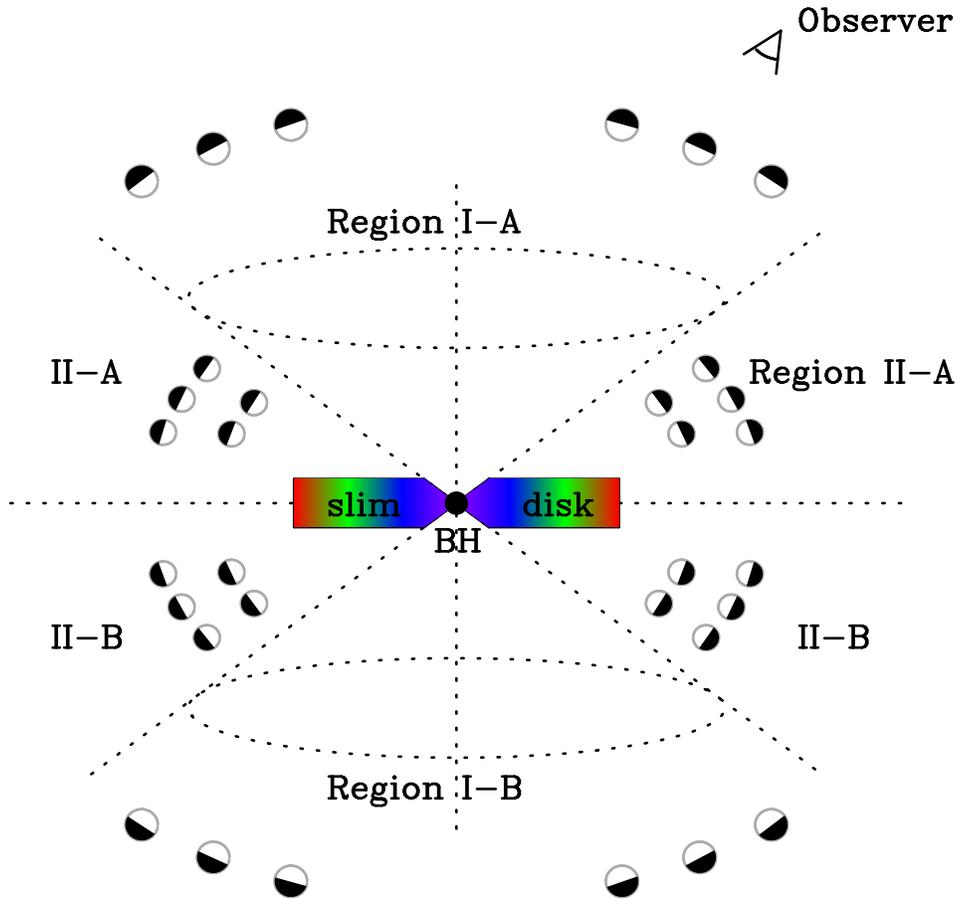}
\end{center}
\caption{\footnotesize A cartoon of the BLR illuminated by a slim disk. 
Regions I and II are divided by the funnel of the disk. The clouds emit
H$\beta$ photons anisotropically. The inward-facing part of the cloud (white) 
is ionized, whereas the outward-facing part of the cloud (black) is not 
ionized.  H$\beta$ photons are only emitted from the inward face. Given 
the anisotropy of H$\beta$ emission, we divide the BLR into Regions I-A 
and I-B and Regions II-A and II-B. The observed H$\beta$ fluxes in these 
regions have different contributions in response to the varying continuum.
Clouds in the BLR are presumed to be distributed homogeneously 
around the accretion disk, but the clouds plotted here are only those
emitting H$\beta$.}
\label{hubble_diagram}
\end{figure*}
\vglue 0.5cm

The ionizing flux for hydrogen atoms is usually defined by 
$F_{\rm ion}=\int_{\rm 13.6eV}^{\infty} F_{\nu} d\nu$. Since H$\beta$
emissions are mainly driven by soft X-rays (e.g., Kwan \& Krolik 1979), 
we also define $F_{\rm SX}=\int_{\rm 0.1keV}^{\infty}F_{\nu} d\nu$.
Figure 5 shows the dependence of $F_{\rm ion}$ and $F_{\rm SX}$ 
on self-shadowing effects. Models are calculated 
for slim disks with different accretion rates $\mathdotM$. We find that SS 
disks have much weaker anisotropy of their radiation fields, which can be 
roughly approximated by $\cos\theta$, and hence self-shadowing effects can 
be neglected.  Moreover, the ionizing photons tend to saturated with increasing 
$\mathdotM$ and have strong anisotropy when $\theta\gtrsim \Theta_{\rm fun}$, 
which leads to anisotropic illumination of the BLR.  Lastly, the anisotropy 
within the funnels of the slim disk varies with $\cos\theta$, as is the case
in thin disks. This can be understood by the fact that the projected area of 
the funnels is viewed by a factor $\cos\theta$.

For slim disks, the strong anisotropy of the radiation field divides the 
BLR into two regions marked as Regions I and II in Figure 3. Region I is 
unshadowed and Region II is shadowed.  For slim 
disks with $\mathdotM=500$, the ionizing flux ratio 
$F_{\rm I}/F_{\rm II}\sim 5-14$ for $\theta=10^{\circ}-40^{\circ}$ and 
$\theta=60^{\circ}-70^{\circ}$. Obviously, this 
ratio depends on $\mathdotM$ because the self-shadowed regions rely on the 
funnel opening angle determined by $\mathdotM$. The strong self-shadowing 
effects on the illumination of the BLR govern diverse observed appearance of 
the BLR, in particular the H$\beta$ reverberation with varying continuum.

In this section, we show that the BLR is divided into two regions (I and II)
in light of the radiation anisotropy of slim disks. We will show that the
clouds in the two regions reverberate differently in response to a varying 
continuum.
We stress the natural appearance of two distinct regions due to the anisotropic 
radiation from slim disks. In this work, we neglect the reflection of the
photons in the funnel of the slim disk. This additional source of heating for 
the funnel could increase the funnel effects in slim disks.  We also do not 
consider general relativistic effects in the disk structure or photon 
propagation. All these effects will enhance the self-shadowing effects. 
The present results should be conservative considerations.

\section{Observational Consequences}
One of the main goals of the present paper is to show the effects of the 
anisotropic disk radiation on the BLR in systems with high accretion rates. 
It is beyond the scope of the present paper to systematically build up 
a complete model of the BLR; however, we will outline the major 
consequences of the anisotropic illumination produced by slim disks. Our 
present discussion has the greatest bearing on NLS1s.

In order to investigate the observational consequences on the BLR, 
we have to reasonably postulate some basic physical properties and spatial 
distribution of the clouds.  We assume that a large number of clouds occupy
the entire space surrounding the accretion disk. Considering the anisotropic 
illumination of the clouds, their spatial distribution falls into two regions. 
The physical properties of the clouds are based on the calculations of Ferland 
et al. (2009).  For clouds with large column density, we simplify the 
cloud's emission into two parts: (1) the front surface facing the ionizing 
source and (2) the back facing away from the ionizing source. The former 
is called inward emission, and the latter is referred to as outward emission. 
The ratio of inward to outward emission depends on the column density of the 
clouds.

The photoionization calculations of Ferland et al. (2009) show that, for a
typical column density of $N_{\rm H}\gtrsim 10^{22}{\rm cm^{-2}}$, the outward
component of the line intensity is characterized by
$I$(\feii)/$I$(H$\beta)_{\rm tot}\sim 10$ and 
$I$(H$\beta$)$_{\rm out}$/$I$(H$\beta)_{\rm tot}\sim 0.1$,
where $I({\rm H\beta})_{\rm tot}$ is the sum of the inward and outward 
emission.  This indicates that optical 
\feii\, emission is isotropic whereas H$\beta$ 
emission is anisotropic (Ferland et al.  2009). It is reasonable to assume 
that the clouds in AGNs with high accretion rates may have large $\NH$.
Otherwise, the clouds will be blown away by the super-Eddington luminosity. 
Additionally, the reverberation response of H$\beta$ to the varying continuum 
indicates that the BLR clouds are not fully ionized.  We assume that clouds 
with large column density ($>10^{22}{\rm cm^{-2}}$) in Regions I and II emit 
H$\beta$ anisotropically but that \feii\, is emitted isotropically. 
However, but we bear in mind that the column density of the clouds is 
highly uncertain; we will briefly discuss the effects on low-$\NH$ clouds.

\subsection{Two Dynamical Regions of the BLR}
If the BLR has a constant product $n_eU$, we naively expect the size of 
the BLR to scale with the ionizing luminosity as 
$\rblr\propto L_{\rm ion}^{1/2}$, where $L_{\rm ion}$ is the ionizing 
luminosity. This size scaling agrees surprisingly well 
with the reverberation mapping data for H$\beta$ (Kaspi et al. 2000;
Bentz et al. 2013). This luminosity-size relation can be used to constrain the 
two regions of the BLR affected by the anisotropic illumination (see Figure 3). 
Considering the high anisotropy of the radiation from slim disks, the size 
scales of the two BLR regions are related through
\begin{equation}
\frac{\rblrI}{\rblrII}=\left(\frac{L_{\rm ion,I}}{L_{\rm ion,II}}\right)^{1/2}
                            =\left(\frac{\fionI}{\fionII}\right)^{1/2},
\end{equation}
where $F_{\rm ion,I}$ and $F_{\rm ion,II}$ are the fluxes\footnote{ 
Soft X-rays are the main driver for H$\beta$ emission, as pointed out by the
referee. In the subsequent 
discussion, $F_{\rm ion}$ represents $F_{\rm SX}$ for H$\beta$ emission; 
however, we retain $F_{\rm ion}$ as a general symbol.}
at the typical opening angles 
at radius $r_0=10^4\rg$. As shown in Figure 3, we have the ratio of 
$\rblrI/\rblrII=(2.0,2.5,3.5,4.6)$ for $\mathdotM=(50,100,500,1000)$ and 
$(\theta_{\rm I},\theta_{\rm II})=(20^{\circ},70^{\circ})$, where 
$\theta_{\rm I,II}$ are the orientation angles in Regions I and II. From 
these calculations, we approximate
\begin{equation}
\frac{\rblrI}{\rblrII}\approx 2.0\mathdotM_{50}^{0.3},
\end{equation}
where $\mathdotM_{50}=\mathdotM/50$. We point out that the size of the
BLR ($\rblr$) used here is the ionization front of the BLR. In the presence of 
self-shadowing effects, the ionization front shrinks rather than changes the 
spatial distributions of clouds. Figure 6 gives a cartoon of the two-region 
BLR, showing how the H$\beta$-emitting clouds in the two regions are located 
at different distances from the center. 

Since the clouds in Regions I and II have different distances from the center,
the velocity widths of H$\beta$ from the two regions should also be different. 
For lines broadened by the black hole potential, we have
\begin{equation}
\frac{V_{_{\rm H\beta, I}}}{V_{_{\rm H\beta, II}}}
       =\left(\frac{\rblrI}{\rblrII}\right)^{-1/2}
        \approx 0.7\mathdotM_{50}^{-0.15},
\end{equation}
where $V$ is the full width half-maximum of the H$\beta$ profile.  The clouds 
in the two regions have different kinematics, such that
$V_{_{\rm H\beta, I}}<V_{_{\rm H\beta, II}}$.  This implies that the observed 
H$\beta$ emission is composed of two independent components: a very broad 
component (VBC) with $V_{_{\rm H\beta,II}}$ and an intermediate-breadth component 
(IBC) with $V_{_{\rm H\beta,I}}$. The VBC and IBC are dynamically 
independent. The strong anisotropic radiation indicates that the ratio of the 
two region sizes, and hence also the ratio of their respective velocity 
widths, depends on accretion rate.  

As illustrated in Figure 6,  Regions I and II have altogether four subsections 
(see the figure caption for a detailed explanation).  The separation between 
Regions II-A and II-B are caused by the complete shadowing of the slim disk. 
The total H$\beta$ emission comes from the four subsections, depending on 
the observer's orientation with respect to the funnel. We stress that all 
four subsections contribute to the BLR emission. Real BLRs may be more 
complicated than the schematic envisioned here, but the main features and 
consequences of anisotropic illumination of the BLR should be robust.

As H$\beta$ arises from photoionized clouds, its flux can be written 
as $F_{\rm H\beta}=C\eta_{_{\rm H\beta}}F_{\rm ion}$, where $C$ is the 
covering factor of the BLR and $\eta_{_{\rm H\beta}}$ is the efficiency 
for reprocessing ionizing photons into H$\beta$. For simplicity, we 
assume that the radiation efficiency $\eta_{_{\rm H\beta}}$ is the same 
for all clouds. The total observed flux from the four regions is given by
\begin{equation}
F_{\rm H\beta}^{\rm tot}
       =\left(\xiIA+\xiIB\right)\CI\eta_{_{\rm H\beta}}\fionI+
        \left(\xiIIA+\xiIIB\right)\CII\eta_{_{\rm H\beta}}\fionII,
\end{equation}
where $\xi$ is the anisotropic factor for H$\beta$ emission, and its subscript 
refers to the corresponding subsection. These factors depend on the orientation of 
the observer as well as on the column density of the clouds.  For a pole-on 
observer, we have the characteristic values $\xiIA=0$, $\xiIB=1$, and $\xiIIA=\xiIIB=1/2$.

If the clouds are distributed homogeneously around the central ionizing source, 
the covering factor $\CI\propto \Delta\Omega=2\pi (1-\cos\Theta_{\rm fun})$
and $\CII\propto 2\pi \cos\Theta_{\rm fun}$, where $\Delta\Omega$ is the 
solid angle subtended by the funnel. The flux ratio of VBC and IBC for
H$\beta$ depends on the covering factors, such that
\begin{equation}
\frac{\fhbetaI}{\fhbetaII}
 =\frac{\left(\xiIA+\xiIB\right)\CI\fionI}{\left(\xiIIA+\xiIIB\right)\CII\fionII}     
 =3.6\frac{\left(\xiIA+\xiIB\right)(1-\cos\Theta_{\rm fun})}
     {\left(\xiIIA+\xiIIB\right)\cos\Theta_{\rm fun}}\mathdotM_{50}^{0.6}
\end{equation}
for $(\theta_{_{\rm I}},\theta_{_{\rm II}})=(20^{\circ},70^{\circ})$.
For pole-on observers, we have 
$\fhbetaI/\fhbetaII\approx (\cos^{-1}\Theta_{\rm fun}-1)\mathdotM_{50}^{0.6}
\approx 0.4\mathdotM_{50}^{0.6}$, 
which is a function of $\mathdotM$. Future tests can be done by 
evaluating these potential empirical correlations.

For AGNs in the SS regime, self-shadowing is very weak so that the boundary
between Regions I and II disappears.  Apart from the factor
$\cos\theta(1+2\cos\theta)$, the effects of anisotropic radiation can be
neglected.  All the properties discussed in this paper reduce to those of
normal (sub-Eddington) AGNs. The two components of the BLR (VBC and IBC)
merge together into an approximately single Gaussian profile, consistent with
observations of most quasars and Seyfert 1 galaxies.

Note that the condition of constant $n_eU$ in the BLR may not be a unique 
explanation for the $R_{\rm BLR}-L$ relation. In the locally optimally 
emitting cloud model of Baldwin et al. (1995), the ionization parameter 
as well as the BLR density have a large range of values.   
An alternative view is that the $R_{\rm BLR}-L$ 
relation may result from the effects of dust sublimation in the BLR (Laor \& 
Draine 1993; Netzer \& Laor 1993).  It is not the goal for the present paper 
to enter into this debate.  We simply employ the $R_{\rm BLR}-L$ relation to 
discuss the ionization front of the BLR and how it is anisotropically 
illuminated by a slim disk.  Our results do not depend on the physical 
interpretation of the  $R_{\rm BLR}-L$ relation.

\subsection{H$\beta$ Reverberation}
Obviously, the manner in which H$\beta$ responds to the varying continuum 
depends on the properties of the clouds in the four subregions.  The actual 
observed reverberation is an averaged response by the spatial distribution 
of clouds (or averaged emissivity of the spatial distribution of clouds). We 
discuss here the characteristic time lags of the BLR. Decomposing the VBC 
and IBC of H$\beta$, we find that the two components should have 
different lags. Consider that the H$\beta$ emission from 
clouds in Region I-A is invisible while that from Region I-B 
is visible and reverberates with the varying continuum. Reverberation of 
H$\beta$ has a lag of $\rblrI(1+\cos\pthetaI)/c$ with the continuum, where
$\pthetaI$ is the mean orientation angle of clouds in Region I. 
Following Kaspi et al. (2000), we assume $\tau_0=\rblrI/c$, and, hence,
for a pole-on observer,
\begin{equation}
\tauI=(1+\cos\pthetaI)\tau_0\approx 2\tau_0,
\end{equation}
which, for small $\pthetaI$, is significantly longer than $\tau_0$.
This longer lag is caused jointly by the anisotropic emission of H$\beta$ 
and the funnel effect. This is very significant for observational 
tests. 

Clouds in Region II are located at low altitudes. Although H$\beta$ emission 
is anisotropic, the inward emission of H$\beta$ from most clouds will 
be visible. The time lag of H$\beta$ from Region II-B can be approximated
by $\rblrII(1+\cos \pthetaII)/c$, where $\pthetaII$ is the average orientation
angle of clouds in Region II.  Clouds in Region II-A 
are visible.  The H$\beta$ lags from Region II-A cannot be longer than 
$\rblrII(1-\cos\pthetaII)/c$, while the clouds in Region II-B cannot be 
shorter than $\rblrII(1+\cos\pthetaII)/c$.  This indicates that the VBC has 
two lags in response to continuum variability, implying that the 
cross correlation function of the VBC should have two peaks.  The ratio of the 
two peaks measured by a pole-on observer is given by
\begin{equation}
\frac{\tau_{_{\rm H\beta,IIA}}}{\tau_{_{\rm H\beta,IIB}}}
              =\frac{1+\cos\pthetaII}{1-\cos\pthetaII},
\end{equation}
in which $\pthetaII$ is a function of $\mathdotM$. For the extreme case in 
which $\pthetaII\approx 3\pi/8$ (i.e., the average angle between $\pi/4$ and $\pi/2$), 
we have $\tau_{_{\rm H\beta,IIA}}/\tau_{_{\rm H\beta,IIB}}\approx 2.2$. This results
purely from the effects of the anisotropic illumination of slim disks. When 
the illumination is homogeneous, the two peaks will merge. Considering 
the symmetric distribution of clouds in Regions II-A and II-B, 
the ratio is independent of the covering factor of the regions and is only 
sensitive to the orientation angles of Region II. This sensitivity provides
an opportunity to observationally {\it measure}\ the orientation angle. This 
suggested effect is based on the fact that clouds in Region II have a 
large column density to reverberate with the varying continuum.

Comparing the time lags between Regions I and II, we have the following ratios:
\begin{equation}
\frac{\tauI}{\tauII}=\left\{\begin{array}{l}
\displaystyle\frac{1+\cos\pthetaI}{1+\cos\pthetaII}
\left(\frac{\rblrI}{\rblrII}\right)\approx 3\mathdotM_{50}^{0.3},\\
   \\
\displaystyle\frac{1+\cos\pthetaI}{1-\cos\pthetaII}
\left(\frac{\rblrI}{\rblrII}\right)\approx 6\mathdotM_{50}^{0.3},
\end{array}\right.
\end{equation}
for $\pthetaI=20^{\circ}$ and $\pthetaII=70^{\circ}$. Equation (22)
includes two kinds of reverberations described by Equation (20) and 
(21). It is very important that the ratios depend on the dimensionless 
accretion rate $\mathdotM$. We note that the radiated luminosity 
tends to be saturated for SEAMBHs, and the SEDs will be insensitive to 
$\mathdotM$. This makes it very difficult to constrain
the true value of $\mathdotM$. Fortunately, the ratio of fluxes and time 
lags of H$\beta$ emission are mildly sensitive to $\mathdotM$, and so we 
may be able to constrain $\mathdotM$ from monitoring the two components of 
H$\beta$ emission.  Equation (22) provides a useful way to estimate 
$\mathdotM$, through the numerical approximation
 
\begin{equation}
\ln \mathdotM_{50}=3\ln(\tauI/\tauII)-4.6. 
\end{equation}

The above discussion assumes that the ionized clouds in Region I emit 
anisotropically, which may or may not strictly hold in all cases.  
Nevertheless, we expect that on average the clouds in Region I will have 
longer lags than those in Region II.  We further assumed that the clouds in 
Region II are not fully ionized.  If the ionization parameter in Region II is 
much higher than that in Region I, H$\beta$ from Region II may not 
reverberate.  In such a case, while both VBC and IBC are present in the 
composite line profile, only the IBC will respond to variations in the 
ionizing continuum.  

\subsection{The $\taublr-L$ Relation}
The $\rblr-L$ relation (Kaspi et al. 2000;
Bentz et al. 2013) is roughly consistent with the BLR having a 
constant $n_eU$.  The observed tightness of the relation, 
strongly suggests that the ionizing continuum received by the BLR clouds is 
the same as that seen by the observer.  Indeed, most of the $\sim$50 
reverberation-mapped AGNs are sub-Eddington systems, whose SS disk is not 
expected to produce strong anisotropic emission.  Super-Eddington AGNs, 
however, being strongly affected by the self-shadowing effects discussed in 
this work, produce broad-line emission from two distinctive regions, each with 
a different lag.  The observed lag is the sum of the lags from Regions I and II.

If the H$\beta$ emission is not decomposed into the VBC and IBC,
the lag of the total H$\beta$ flux $F_{\rm H\beta}^{\rm tot}$ can be
determined by the cross correlation function,  
\begin{equation}{\begin{array}{lll}
{\rm CCF}&=&F_{\rm H\beta}^{\rm tot}\otimes \fionI \\
         & &  \\
         &\propto&
          \left(\xiIA+\xiIB\right)\CI f_1(t,\tau_{_{\rm I}})+
          \left(\xiIIA+\xiIIB\right)\CII qf_2(t,\tau_{_{\rm II}}),
          \end{array} }
\end{equation}
where $q=\fionI/\fionII$ is the ratio of fluxes in the two regions,
$\fhbetaI(t)=\CI\eta_{_{\rm H\beta}}\fionI(t-\tau_{_{\rm I}}), $
$\fhbetaII(t)=\CII q\eta_{_{\rm H\beta}}\fionI(t-\tau_{_{\rm II}}),$
$f_1(t,\tau_{_{\rm I}})=\fionI \otimes \fionI(t-\tau_{_{\rm I}})$ 
and $f_2(t,\tau_{_{\rm II}})=\fionI\otimes \fionI(t-\tau_{_{\rm II}})$ 
are the autocorrelation functions with peaks at $\tau_{_{\rm I}}$ and 
$\tau_{_{\rm II}}$, which are the lags of the two components. Here we note
that the observed continuum only comes from Region 
I for observers with small $\theta$. The observed lag is the averaged 
values of $\tau_{_{\rm I}}$ and $\tau_{_{\rm II}}$, which are governed
by the covering factors of Regions I and II and the anisotropy ($\xi$) of 
H$\beta$ emission from the BLR clouds. Since the covering factors are 
related to the fluxes of the two components and $\mathdotM$ (see eq. 8), 
for extremely high accretion rates $\CII$ may dominate $\CI$, in which case
$\tau_{_{\rm II}}$ can be significantly shorter than the lag predicted from the
$\taublr-L$ relation. Therefore, we predict that super-Eddington sources should 
significantly deviate from the normal $\taublr-L$ relation.

The ratios of fluxes and lags of the two components depend on the funnel 
opening angle and accretion rate. Future reverberation mapping observations 
can be analyzed in detailed to constrain the funnel opening angle and 
accretion rate. This can be achieved in practice for actual reverberation 
mapping data. Note that the timescale of Keplerian rotation is 
given by $t_{\rm Kep}\approx 10~m_7r_4^{3/2}$ yr, where $r_4=r/10^4\rg$. 
This timescale is much longer than typical
monitoring periods (about 6 months), and thus the individual regions should
retain stable emission profiles during the course of an observational campaign.

Current models do not provide strong constraints on the BLR covering factor.
The relationship between the SED of an AGN and the properties of the BLR
clouds has been discussed by Fabian et al. (1986) and Bechtold et al. (1987)
in the context of the two-phase model of quasar emission-line regions
(Krolik et al. 1981).  The preliminary discussion on anisotropic effects by
Bechtold et al. (1987), interestingly, is based on the NLS1 PG 1211+143.
Goad \& Wanders (1996) investigate some effects of anisotropic illuminations 
on the BLR clouds. It has been observationally well established that accretion 
disks with higher accretion rates have
fainter hard X-rays than ones with lower rates because of Comptonization
of the hot corona by efficient cooling (e.g., Wang et al. 2004; Kelly et al. 
2008; Cao 2009; Uzdensky 2013; see also Czerny et al. 2003; Proga 2005). 
This property holds in AGNs and quasars,
from low to high redshifts (e.g., Shemmer et al. 2008; Brightman
et al. 2013; Fanali et al. 2013). In such a case when the ionizing spectrum is
very soft, the lower Compton temperature and enhanced cooling may actually
inhibit the existence/formation of clouds within the funnel (Krolik et al.
1981). The covering factor ($\CI$ and $\CII$) and its variation with angle
from the disk axis should be obtained self-consistently from models that
employ the SEDs of slim disks.

\subsection{Reverberation of \feii\, Emission}
AGNs with high accretion rates are known to have strong iron (\feii) lines in 
the optical and UV bands (e.g., Boroson 2002).  A systematic study of Sloan 
Digital Survey (SDSS) quasars shows that \feii\, emission has FWHM and 
velocity offsets similar 
to the intermediate-width component of H$\beta$ (Hu et al. 2008a,b, 2012). This 
velocity offset anti-correlates with the Eddington ratio, so that it vanishes 
for most NLS1s.  Generally, however, it has been difficult to detect 
variability in \feii\, emission in AGNs.  Barth et al. (2013) recently reported
\feii\, variability and a successful determination of its time lag in two AGNs; the 
\feii\ lag is longer than that of H$\beta$ by a factor of $\sim 2$. With the goal 
of studying the properties of SEAMBHs, Du et al. (2014) and Wang et al.
(2014) have been carrying out a large campaign to monitor AGNs and quasars with 
high accretion rates. Detailed analysis of \feii\, will be carried out
in a separate paper (C. Hu et al. 2014, in preparation), but here we note that 
the longer lags expected for \feii\ are generally consistent with the 
predictions of our simple model.  For clouds with large $\NH$, 
\feii\,$\lambda$4558 \AA\, emission is isotropic and the observed \feii\, 
strength increases with $\mathdotM$ (see Figure 3 in Ferland et al. 2009).
However, the UV complex \feii\, $\lambda$2445 \AA\,
emits predominantly inward due to its large optical depth.  We 
expect \feii\,$\lambda2445$ \AA\, to reverberate with a longer lag than 
\feii\,$\lambda$4558 \AA.
This may explain why \feii\,$\lambda4558$ \AA\, is strong whereas 
\feii\,$\lambda2445$ \AA\, is expected to reverberate later than 
\feii\,$\lambda$4558 \AA\, with the continuum. 
The different behavior of \feii\ in the optical and UV bands can be 
tested observationally using simultaneous observations.

As we argued in the last paragraph of \S3.1, it is possible that Region II has 
high $U$. A high $U$ in Region II suppresses \feii\, emission relative to 
Region I. This prediction of our model indicates that \feii\ variations follow 
the IBC and reverberate with the varying continuum. To date, only two AGNs 
have been successfully monitored for \feii\ variations (Barth et al. 2013), 
and neither is powered by a slim disk. A key test of our present model is to 
monitor the \feii\, variations and check if they follow the varying continuum.

\subsection{BLR and Outflows}
The scaled distances of the BLR ionization fronts are 
based on the paradigm of a constant product $n_eU$. However, this has
additional effects on the clouds\footnote{The existence of clouds in the
two-phase BLR model (Krolik et al. 1981) is often questioned because of the expected 
short lifetimes of the clouds due to dynamical friction in the hot medium (Mathews \& 
Ferland 1987). Non-steady models of the BLR avoid this difficulty because the number 
of clouds is set through the balance between cloud production and 
cloud destruction in the hot medium (Czerny \& Hryniewicz 2011; Wang et al. 2012).}. 
The self-shadowing effects may 
lower the gas pressure of the surrounding medium of the clouds in the shadowed regions 
so that the internal pressure causes the clouds to expand from $r_{\rm cl}^0$ 
to $r_{\rm cl}$. For mass conservation to hold, the column density of clouds 
in the shadowed regions changes as $\NH=\NH^0(r_{\rm cl}^0/r_{\rm cl})^2$, 
where $\NH^0$ is the column density of the unshadowed regions. 
The column density of the clouds in Region II is lower than that in Region I
so that the clouds in Region II may transition from $\tau\gg1$ to $\tau\sim 1$;
this speculation needs to be verified with future detailed calculations. The case of
$\tau \gg1$ clouds in Region II has been discussed in previous sections.  What about 
the potential situation of $\tau\lesssim 1$ clouds?  Each cloud may be
mostly ionized, and the H$\beta$ flux depends on the reprocessing efficiency of 
the clouds.
More interesting, clouds in Region II may undergo acceleration driven by line radiation 
pressure with a large force multiplier to 
produce outflows  from this region. Observationally, the VBC is expected to show 
blueshifts and reverberate little, if at all, with the continuum. 

If clouds are blown away in Region II, how are they replenished?  According to 
numerical simulations, slim accretion disks develop strong outflows 
(Ohsuga et al. 2005;  McKinney et al. 2014; Sadowski et al. 2014), which may be
the very source of clouds for the BLR (Murray \& Chiang 1997; Kashi et al. 
2013). This physical picture in principle can be tested directly through 
velocity-resolved (two-dimensional) reverberation mapping, which can help to 
diagnose the dynamics of the BLR (Denney et al. 2009, 2010; Grier et al. 
2013). However, as we argue below, velocity-resolved reverberation mapping
technique is still limited because it is not able to uniquely map the
velocity field in space. For AGNs with extremely high accretion 
rates, fortunately, self-shadowing effects provide an opportunity 
to probe the connection between outflows and the BLR through the 
different reverberation properties of clouds in Regions I and II, which 
can be decomposed as separate components of the H$\beta$ line. 

\def\rvbc{R_{_{\rm VBC}}}
\def\cvbc{C_{_{\rm VBC}}}

In principle, we can estimate the outflow rates from the disk surface if we 
can detect shifts ($V_{\rm out}$) of the VBC (Region II) from reverberation 
mapping observations. The radius of the outflow can be estimated by the size 
of the VBC region, $\rblrII$, and hence the mass outflow rate 
\begin{equation}
\dot{M}_{\rm out}=2\pi \rblrII^2V_{\rm out}\CII\rho_{\rm c}, 
\end{equation}
where $\rho_{\rm c}$ is the mass density of the emitting clouds. In principle,
$\CII$ can be determined from the equivalent width of the VBC, and 
$\rho_{\rm c}$ can be estimated from photoionization calculations. 
If the outflow is composed of discrete clouds, $\CII\rho_{\rm c}$ 
is the average density of the outflow. Previous 
attempts to obtain mass outflow rates in AGNs, for example from UV
absorbers (e.g., Arav et al. 2013), have been frustrated mainly by the lack of 
reliable distance estimates.  The method proposed here solves this problem 
because $\rvbc$ can be determined accurately by reverberation mapping.  Comparing 
$\dot{M}_{\rm out}$ with $\dot{M}_{\bullet}$, we can set rigorous constraints 
on the theoretical model of slim disks from observations.

We bear in mind that clouds in Region I may be fully ionized and have no 
reverberation with the varying continuum. In that case, an outflow would 
develop within the solid angle of the funnel. Needless to say, improved 
velocity-resolved reverberation mapping
should be applied to test whether the IBC shows any evidence for velocity 
shifts.

\subsection{The Lorentzian Profile of H$\beta$}
Apart from the different reverberation response of H$\beta$ from Regions I 
and II expected in the time domain, the single-epoch spectra of AGNs powered 
by slim disks may exhibit kinematic signatures that distinguish them from 
AGNs with lower accretion rates. The H$\beta$ profiles of NLS1s are close to 
Lorentzian (e.g., Zhou et al. 2006), but the reason for this is 
insufficiently understood. The Lorentzian profile 
may depend on the rotation, inflows, outflows, turbulence, and even 
the orientation of the BLR to the observer (see Kollatschny \& Zetl 2011). 
A Lorentzian profile in principle can be decomposed mathematically 
into two distinct Gaussian components, which plausibly may be 
related to the two regions invoked in our scenario.  The highly 
anisotropic illumination of slim disks naturally predicts this type of 
two-component structure of the BLR.  Reverberation mapping provides a 
promising method to decouple these two components in the time domain.

Velocity-resolved reverberation mapping, which identifies the lags of 
different parts of the line profile with the varying continuum, has 
been applied to infer the dynamical structure of the BLR 
(e.g., Denney et al. 2009, 2010). This idea generally works 
if the velocity field of the BLR clouds is not degenerate 
with respect to their spatial distribution. However, the velocity field 
does not uniquely map to the spatial domain because the observer only views
the velocity projected along the line of sight.  For example, the core 
of the H$\beta$ profile may come from different regions of the 
BLR. This limits the application of velocity-resolved reverberation mapping to 
reconstruct the BLR, in particular for high-accretion rate AGNs. 
By contrast, the decomposition of the H$\beta$ emission into two components 
allows us to get time lags for the two different regions of the BLR.  Current 
Markov Chain Monte Carlo analysis of AGN light curves employed to 
estimate black hole masses and BLR dynamics assumes that the BLR is homogeneous 
(Pancoast et al. 2011; Li et al. 2013). Future improvements of this technique 
should include the inhomogeneous illumination of the BLR.

\section{Discussion}
\subsection{Narrow-line Region}
The \oiii\, $\lambda$5007 emission line is a prominent feature in the spectra 
of AGNs and quasars (e.g., Vanden Berk et al. 2001), but it is generally 
relatively weak in NLS1s (Osterbrock \& Pogge 1985; Boller et al. 1996) and 
especially in systems with high accretion rates (Boroson 2002; Shen \& Ho 
2014).  The inverse correlation between the strength of the narrow lines and 
Eddington ratio holds not only for \oiii\ but also for other high-ionization 
narrow lines (Shen \& Ho 2014).  The fundamental reason why the amount of 
narrow-line emission is connected with the accretion rate is still unclear.

Here we suggest that the suppression of high-ionization narrow-line emission 
with increasing Eddington ratio arises naturally from the strong anisotropy
of the ionizing radiation field of slim disks.  The deep funnel of the disk 
confines the ionization cone to illuminate only a portion of the narrow-line 
region, thus reducing the strength of the narrow lines, especially those of 
high ionization. This explanation agrees with the interpretation 
of ``Eigenvector 1,'' as suggested by Boroson \& Green (1992).
Usually the ionization cone of \oiii\, is thought to be 
determined by the dusty torus (e.g., Schmitt et al. 2003; Bennert et al. 
2006). Here we propose that the collimation begins on much smaller scales, 
and that it is fundamentally connected with the structure of slim disks.  
This idea can be tested by measuring the opening angle of ionization cones 
in AGNs with well-resolved narrow-line regions\footnote{Since most 
Seyfert 1 galaxies are relatively face-on sources, projection effects will 
be significant. \oiii\, cones are observed more
often in Seyfert 2 galaxies. Only a few of NLS1s (Akn 564, Mrk 493, Mrk 1044,
NGC 4051, NGC 5506, NGC 7469, and Mrk 766) have
\oiii\, images (Schmitt et al. 2003; Fischer et al. 2013), so that it is 
currently difficult to make a statistical test.}.

Additionally, the \oiii\, profile of NLS1s often has blue wings (e.g., 
Boroson 2005).  With Eddington ratios of $\sim 2$ (equation 2), we would like 
to point out that slim disks may generate radiation pressure-driven outflows 
on large scales (e.g., King et al. 2011; Pounds \& King 2013).  Large-scale 
outflows driven by slim disks may also explain the correlation 
between extended radio emission and blue-asymmetric \oiii\, lines
seen in obscured luminous quasars (e.g., Stocke et al. 1992; Zakamska \& 
Greene 2014).

\subsection{Weak-line Quasars}
Some quasars with nearly featureless spectra have been found in SDSS (e.g., 
Plotkin et al. 2010); their physical nature is not understood. X-ray 
observations of radio-quiet weak-line quasars show that they have steeper 
spectra in the 2-10 keV band and an order of magnitude fainter X-ray emission 
than normal quasars (Wu et al. 2011, 2012).  Empirically it has been established
that the steepness of the X-ray slope correlates with Eddington ratio (e.g., 
Lu \& Yu 1999; Wang et al. 2004). This can be understood phenomenologically by 
appealing to more efficient Compton cooling of the hot corona by seed photons 
from the cold part of the disk in systems with high accretion rates.  This not 
only steepens the X-ray spectrum but also reduces the output of hard X-rays.
Slim disks may naturally suppress thermal instabilities and prevent the 
formation of a two-phase medium, thus suppressing strong line emission.
Furthermore, self-shadowing obscures the X-rays from the hot 
corona of the disk surface and enhances the suppression of the instabilities. 
These two factors reduce the strength of the line emission and may explain the 
appearance of weak-line quasars. As a specific example, the 
narrow-line quasar PHL 1811, which has X-ray spectral properties very similar 
to those of NLS1s, has a very soft SED and shows relatively weak emission lines 
in the optical and UV bands (Leighly et al. 2007).  We attribute the weak line 
emission in PHL 1811 to the self-shadowing effects of its slim disk. Alternatively, 
Laor \& Davis (2011) suggest that the suppressed line emission of weak-line quasars 
can be explained by the low temperature of the accretion disks around very massive 
black holes.  This hypothesis can be tested with detailed multiwavelength 
characterization of the SEDs and accurate black hole mass measurements.

\subsection{Near-infrared Reverberation with the Varying UV Continuum}
The well-known relation between reverberation in the near-infrared ($\tauNIR$) 
with the varying optical-UV continuum in AGNs (Suganuma et al. 2006) is 
consistent with the sublimation radius of dust particles evaporated by 
a central UV source, 
\begin{equation}
r_{\rm torus}=0.13~ L_{44}^{1/2}T_{1800}^{-2.8}~{\rm pc}, 
\end{equation}
where $L_{44}=L_{\rm UV}/10^{44}\,\ergs$ is the UV luminosity of the central
energy source and $T_{1800}=T/1800$ K is the evaporation temperature 
of dust particles (Barvainis 1987). This reverberation relation results
from the reprocessed emission of the UV radiation from the central disk. It is 
interesting to note that it also sets up strong constraints on the BLR 
(Netzer \& Laor 1993). However, the self-shadowing effects of slim disks, if 
the opening angle of the funnel is smaller than the opening angle of the 
torus, reduces the heating of the inner edge of the torus, leading to a 
general shrinking of its inner radius.  This potential effect has been noted 
by Kawaguchi (2013) but has yet to be observed. It would be interesting to 
systematically study the near-infrared reverberation to the optical and UV 
emission in AGNs with high accretion rates. If the above picture holds, the 
time lags in the near-infrared will be much shorter than those predicted by the 
$\tauNIR-L_{\rm UV}$ relation of Suganuma et al. (2006). 

\subsection{Dust-poor or Dust-free Quasars}
Hot dust particles are a generic constituent of the dusty torus invoked in AGN 
unification schemes (Antonucci 1993), and they should emit strong infrared 
emission (Pier \& Krolik 1993). This basic expectation is borne out in many 
infrared observations of AGNs (e.g., Neugebauer et al. 1987; Haas et al. 2003;
Hatziminaoglou et al. 2008). However, there is growing evidence for a minority 
of quasars that appear to be dust-poor (Hao et al. 2011) or even dust-free 
(Jiang et al. 2010).  Are these systems really deficient in dust?

The self-shadowing effects from slim disks may provide a simple and natural 
explanation for this class of objects. Again, if the funnel opening angle is
smaller than that of the dusty torus, the torus will not be effectively 
irradiated by the disk.  Further, if the torus is supported by the radiation 
of the accretion disk (Pier \& Krolik 1992), it may collapse in the presence of 
self-shadowing effects from a slim disk, further reducing the ability for the 
torus to be illuminated by the disk.  The torus will thus be infrared-faint 
and appear to be dust-poor, or even dust-free.

A key observational test of our alternative scenario for the origin of
infrared-weak quasars is to quantify the metallicity of the BLR using the 
UV line diagnostics suggested by Hamann \& Ferland (1999).  If a 
quasar is genuinely dust-poor, we expect it also to be metal-poor, 
to the extent that metal and dust content scale with each other in galaxies 
(e.g., Draine et al. 2007). In our 
scenario, infrared-weak quasars are only apparently dust-poor because the 
UV photons do not intercept the torus due to the geometric 
configuration of slim disks.  They should, in fact, have normal dust content, 
and hence normal metal content.

\section{Conclusions}
Based on the vertically averaged equations of an accretion flow in the 
super-Eddington regime, we show that the structure of slim disks contains 
a sharp inner funnel and a flattened outer region that produces strong
self-shadowing and anisotropy of the radiation field. We calculate the SEDs
received by clouds located at different orientations in 
the BLR and show that the anisotropy of the radiation field increases with 
accretion rate.  This allows us to naturally link the structure of the 
central engine and the BLR in AGNs. We demonstrate that self-shadowing leads to 
anisotropic illumination of the BLR clouds, which leads to diverse 
observational properties of broad emission lines such as H$\beta$ and \feii.
The strong anisotropy of the radiation field naturally divides the BLR into 
two regions, which have kinematically distinct line widths and temporal 
response to continuum variations.  These predictions can be tested with 
reverberation mapping of AGNs with high accretion rates (e.g., NLS1s).

We also extensively discuss other observational consequences of the 
self-shadowing effects of slim disks in AGNs. These include: departures from 
the relationship between BLR size and luminosity; longer time lags for 
reverberation of \feii; the Lorentzian profile of broad H$\beta$ in NLS1s;
suppression of high-ionization narrow emission lines in systems with 
high Eddington ratios; origin of weak-line quasars; and reduction of infrared
emission from the dusty torus, which can explain
apparently dust-poor or dust-free objects.

\acknowledgements  JMW thanks H. Netzer for useful discussions and 
suggestions. The authors are grateful to the second referee for his/her
helpful reports to improve the paper. The authors are grateful to the members, 
in particular, C. Hu and Y.-R. Li, of IHEP AGN 
group for stimulating discussions. J.-Q. Ge is thanked for plotting 
Figures 3 and 6. This research is supported by the Strategic Priority 
Research Program $-$ The Emergence of Cosmological Structures of the 
Chinese Academy of Sciences, Grant No. XDB09000000, 
by NSFC grants NSFC-11173023, NSFC-11133006, and NSFC-11233003, and 
by Israel-China ISF-NSFC grant 83/13. LCH receives support from the 
Kavli Foundation and Peking University.

%\vglue 0.3cm
%{\it Notes added in the proof.} After submissions of the revision of this paper,
%we have completed data analysis of the 2013 campaign of Super-Eddington Accreting Massive 
%Black Holes (SEAMBH2013) project through Lijiang 2.4-m telescope in Yunnan Observatory.
%Results of observational data show that the H$\beta$ lags are significantly shorter 
%than the current $\rblr-L$ relation (Du et al. 2014 in preparation), illustrating 
%that the present scenario of
%self-shadowing effects is consistent with the observational data.

%\vglue 0.5cm

\clearpage

%\vglue 1cm

\appendix

\section{Method to Calculate Occultation}
We use the cylindrical coordinate system shown in Figure 3. 
After some algebraic manipulations, we can show that for a cloud in the BLR located 
at $P(r_0,\pi/2,z_0)$, the unity vector of the surface at any 
point on the disk, the direction of the line $PQ$, and the 
direction of the cloud are  
\begin{equation}
\displaystyle
 \vec{n}=\frac{1}{\sqrt{1+H_r^2}}\left(-H_r\cos\phi,-H_r\sin\phi,1\right),
\end{equation}
\begin{equation}
 \vec{n}_1=\left(\frac{1}{d}\right)(-r\cos\phi,r_0-r\sin\phi,z_0-z),
\end{equation}
and
\begin{equation}
\vec{n}_0=\frac{1}{\sqrt{r_0^2+z_0^2}}(0,r_0,z_0),
\end{equation}
respectively, where $d=\sqrt{r^2-2rr_0\sin\phi+r_0^2+(z_0-z)^2}$.

Since the disk obscures itself, we neglect the emission
from the lower half of the plane and from most of the
inner part of the disk through the space around the region between 
the last stable radius and the Schwartzschild radius. In the Newtonian 
approximation of a slim disk, general relativistic 
bending of the photon trajectory can be neglected.

For a cloud located at ($r_0,\pi/2,z_0$), we first get the tangent 
line from the cloud to the surface, which satisfies 
$\vec{n}\cdot \vec{n}_1=0$:
\begin{equation}  
(r-r_0\sin\phi)H_r+z_0-H=0.
\end{equation}
Here $H$ is a function of $r$. We have a curve 
($r_{\rm t},\phi_{\rm t},H_{\rm t}$), where the subscript ``t" is 
the tangent point.  Extending the tangent line until the intersection 
$R$, we have the coordinates of $R$ at $(r_c,\phi_c,H_c)$ with the 
disk surface. The points ($r_{\rm t},\phi_{\rm t},H_{\rm t})$ and 
$(r_{\rm c},\phi_{\rm c}, H_{\rm c}$) consist of a closed curve 
[$r_s(\phi)$] as the boundary between the obscured and unobscured regions. 
The intersection radius $r_c$ can be between $r_{\rm in}$ and 
$r_{\rm Sch}$; in such a case, we set $r_c=r_{\rm in}$. The flux
received by the cloud is given by
\begin{equation}
F_{\nu}(r_0,z_0)=\int_{0}^{2\pi}d\phi\int_{r_s(\phi)}^{\infty}
                 I_{\nu}(r,\phi,z)(\vecn\cdot\vecnnn)(\vecnnn\cdot\vecnn)
                 \frac{d\mathscr{S}}{d_0^2},
\end{equation}
where
\begin{equation}
d_0=\left[r^2\cos^2\varphi+(r\sin\varphi-r_0)^2+(z-z_0)^2\right]^{1/2},
\end{equation}
is the distance of the cloud to the center.

\end{document}